\documentclass[invmat, numbook, draft]{svjour}
\usepackage{times}
\usepackage{amsfonts}
\usepackage{amssymb,epic,eepic} %
\usepackage{amsmath}
\author{ Igor Bjelakovi\'c \inst{1} \and Tyll Kr\"uger \inst{1,2}\and Rainer Siegmund-Schultze\inst{1}\and
  Arleta Szko\l a \inst{1}}
\institute {Technische Universit\"at Berlin
 \thanks{Fakult\"at II - Mathematik und Naturwissenschaften
 \\Institut f\"ur Mathematik MA 7-2
 \\Stra\ss e des 17. Juni 136
 \\10623 Berlin, Germany}, \email{\{igor, tkrueger, siegmund,
    szkola\}@math.tu-berlin.de} \and Universit\"at Bielefeld
 \thanks{Fakult\"at f\"ur Mathematik
 \\Universit\"atsstr. 25
 \\33619 Bielefeld, Germany}}
 \authorrunning{I. Bjelakovi\'c et al.}
\title{Chained Typical Subspaces - a Quantum Version of
Breiman's Theorem}

\newcommand{\hr}{{\cal H}}

\newcommand{\nn}{{\mathbb N}}

\newcommand{\zz}{{\mathbb Z}}
\newcommand{\eps}{{\varepsilon}}        

\begin{document}

\maketitle

\begin{abstract}
We give an equivalent finitary reformulation of the classical
Shannon-McMillan-Breiman theorem which has an immediate translation to the
case of ergodic quantum lattice systems. This version of a quantum Breiman
theorem can be derived from the proof of the quantum Shannon-McMillan theorem
presented in \cite{bksss1}.
\end{abstract}

\section{Introduction}

In classical probability and information theories the Shannon-McMillan
theorem was proved by Breiman in an almost sure version, saying that with
probability one along a trajectory (message string) the individual
probability decreases exponentially at a rate given by the entropy
rate $h$ of
the ergodic stochastic process (over a finite alphabet). From the point of
view of data compression and coding this means that the message string will
not only belong to a typical set of approximate size $e^{nh}$ at a given
time $n$ with overwhelming probability, as is guaranteed by the original
Shannon-McMillan theorem, but it will eventually be captured in the sequence
of typical sets almost surely. So sequential coding at entropy rate should
be possible (and in fact, the Lempel-Ziv algorithm is such a sequential
coding scheme).

It is natural to ask for a corresponding result in the quantum context. In
the quantum situation the notion of an individual trajectory is at least
problematic. So it is not at all obvious what could be an analogue of
Breiman's result. The approach chosen here is to
give an equivalent reformulation of the classical Breiman result which
allows an immediate translation into the quantum setting. In fact, the
typical subsets according to Breiman's theorem are not isolated objects, one
for each $n$, but comprise bundles of trajectories and hence can be chosen
being chained in the sense that the predecessor subset consists exactly of
the same sequences, only shortened for one letter. This simple observation
leads to a finitary, simple and equivalent statement of Breiman's theorem
avoiding the notion of a trajectory. Substituting typical subsets by typical
subspaces of Hilbert spaces, the property of them to be chained can be expressed easily in
terms of partial traces. It turns out that the proof of the quantum
Shannon-McMillan theorem presented in \cite{bksss1}, which is based to a substantial
part on the earlier work of Hiai and Petz \cite{petz}, yields the necessary tools
to prove that typical subspaces can be chained.

\section{A finitary reformulation of the Shannon-McMillan-Breiman theorem}\label{sect2}

Consider an arbitrary (not necessarily stationary) stochastic process $P$
with a finite set $A$ as state space. So $P$ is a probability measure on $%
[A^{\mathbb{Z}},\frak{A}^{\mathbb{Z}}]$ where $\frak{A}^{\mathbb{Z}}$ is the
$\sigma $-field generated by cylinder sets. Let $h$ be a non-negative real
number. We say that $P$ satisfies the condition (B) (with respect to $h$) if
for $P$-a.e. sequence $(\xi _{n})_{n \in \zz}$ the limit $-\frac{1}{%
n}\log P^{(n)}((\xi _{i})_{i=1}^{n})$ exists and equals $h$. Here $P^{(n)}$
is the marginal of $P$ on the cartesian product $A^{(n)}:=\prod_{i=1}^{n}A$.

The Shannon-McMillan-Breiman theorem asserts that ergodic $P$ satisfy (B) with
$h$ being the entropy rate of the process.

Let $\mathbf{x}=(x_{i})_{i=1}^{n}$ be a finite sequence from $A^{\ast
}=\bigcup_{k \in \nn}A^{(k)}$. We denote by $\mathbf{x}_{\flat }$ the
sequence $(x_{i})_{i=1}^{n-1}$ obtained from $\mathbf{x}$ by omitting the
last symbol.
\begin{definition}
We say that $P$ satisfies condition (B*) with respect to $h$ if \ for each $%
\varepsilon $ there exists a sequence $\{C_{\varepsilon
}^{(n)}\}_{n \in \nn}$ of subsets of $A^{(n)}$, respectively,
and a number $N(\varepsilon )$ such that
\begin{enumerate}
\item $C_{\varepsilon }^{(n)}=(C_{\varepsilon }^{(n+1)})_{\flat }$ \ for $%
n\geq 1$ \label{(1)} \\ (sets are chained)

\item $\;\#C_{\varepsilon }^{(n)}\in (e^{n(h-\varepsilon )},e^{n(h+\varepsilon
)})$ for $n\geq N(\varepsilon )$ \label{(2)} \\ (exponential growth rate)

\item $P^{(n)}(\mathbf{x)}<e^{-n(h-\varepsilon )}$ for $n\geq N(\varepsilon )$
and any $\mathbf{x\in }C_{\varepsilon }^{(n)}$ \label{(3)} \\ (upper semi-AEP)

\item$\;P^{(n)}(C_{\varepsilon }^{(n)})>1-\varepsilon $ \label{(4)} \\(typical
sets).
\end{enumerate}
\end{definition}
Except for the concatenation condition (1) these are known properties of the
typical sets for the case of a stationary ergodic $P$. Condition (3) is a
weakened version of the well-known Asymptotic Equipartition Property (AEP),
particularly it ensures that $h$ is an asymptotic entropy rate in the
general (possibly non-stationary) situation.\\
We obtain the equivalence assertion
\begin{lemma}\label{B=B*}
A probability measure $P$ on $[A^{\mathbb{Z}},\frak{A}^{\mathbb{Z}}]$
satisfies (B) iff it satisfies (B*).
\end{lemma}
The proof is entirely elementary. Observe that (1) describes a tree graph of
one-sided infinite trajectories.\\ \\
\textbf{Proof of Lemma \ref{B=B*}:}

1. Assume that (B) is fulfilled. Let
\begin{eqnarray*}
A_{(M,\varepsilon )}:=\{\mathbf{x}\in A^{\mathbb{Z}%
}|\ P^{(n)}((x_{i})_{i=1}^{n})\in (e^{-n(h+\varepsilon
/2)},e^{-n(h-\varepsilon /2)}), \nonumber \\ \forall n \geq M\}.
\end{eqnarray*}
Obviously (B) implies $P(A_{(M,\varepsilon )})\underset{M\rightarrow \infty
}{\longrightarrow}1$ for fixed $\varepsilon $. So there is some $M(\varepsilon
) $ with $P(A_{(M(\varepsilon ),\varepsilon )})>1-\varepsilon $. Define
\begin{eqnarray*}
C_{\varepsilon }^{(n)}:=\{\mathbf{y}\in A^{(n)}|\ \exists
\mathbf{x}\in A_{(M(\varepsilon ),\varepsilon )}
:\ (y_{i})_{i=1}^{n}=(x_{i})_{i=1}^{n}\},
\end{eqnarray*}
let $k(\varepsilon )=\min \{k\in \mathbb{N}|(1-\varepsilon
)e^{k(h-\varepsilon /2)}>e^{k(h-\varepsilon )}\}$ and set $N(\varepsilon
)=\max \{k(\varepsilon ),M(\varepsilon )\}.$ Then (1)-(4) are easily derived,
taking into account that bounds on probabilities imply bounds on
cardinality. So (B*) is a consequence of (B).

2. Assume (B*) to be satisfied. Then it is easy to see that we find a
chained sequence $\{\overline{C}_{\varepsilon }^{(n)}\}$ and some $\overline{%
N}(\varepsilon )$ which fulfil (1)-(4) and even the full AEP condition

\begin{eqnarray*}
P^{(n)}(\mathbf{x)}\in (e^{-n(h+\varepsilon )},e^{-n(h-\varepsilon
  )}),\quad \forall n\geq
\overline{N}(\varepsilon ),\ \forall\mathbf{x\in }\overline{C}_{\varepsilon }^{(n)}.
\end{eqnarray*}
In fact, define
\begin{eqnarray*}
A_{\varepsilon }(n):=\{\mathbf{x}\in A^{\mathbb{Z}}|\ (x_{i})_{i=1}^{n}\in
C_{\varepsilon }^{(n)}\}
\end{eqnarray*}
and observe that $A_{\varepsilon }(n)\underset{n\rightarrow \infty}
\searrow  A_{\varepsilon }$, where $A_{\varepsilon }$ is
the tree of trajectories associated with $\{C_{\varepsilon }^{(n)}\}$.
Condition (4) implies $P(A_{\varepsilon })\geq 1-\varepsilon $. Now let
\begin{eqnarray*}
\widetilde{A}_{\varepsilon }(n):=\{\mathbf{x}\in A_{\varepsilon
}(n)|\ P^{(n)}((x_{i})_{i=1}^{n})\leq e^{-n(h+2\varepsilon )}\}.
\end{eqnarray*}
By (2) it follows that $P(\widetilde{A}_{\varepsilon }(n))\leq
\#C_{\varepsilon }^{(n)}\cdot e^{-n(h+2\varepsilon )}<e^{-n\varepsilon }$.
The Borel-Cantelli lemma implies now that $P(\bigcup_{n\geq m}\widetilde{A}%
_{\varepsilon }(n))\underset{m\rightarrow \infty}\searrow 0$, so there is some $%
m(\varepsilon )$ with $P(\bigcup_{n\geq m(\varepsilon )}\widetilde{A}%
_{\varepsilon }(n))<\varepsilon $. Let $\overline{N}(\varepsilon ):=\max
\{N(\varepsilon /2),m(\varepsilon /2),k(\varepsilon )\}$ (where $k(\varepsilon
)$ was defined above) and
\begin{eqnarray*}
\overline{C}_{\varepsilon }^{(n)}:=\{(x_{i})_{i=1}^{n}\in C_{\varepsilon
/2}^{(n)}|\ \exists \mathbf{w}\in A_{\varepsilon
/2}:\ (w_{i})_{i=1}^{n}=(x_{i})_{i=1}^{n},\nonumber \\ P^{(k)}((w_{i})_{i=1}^{k})>e^{-k(h+%
\varepsilon )},\ \forall k\geq N(\varepsilon )\}.\nonumber
\end{eqnarray*}
We have, denoting by $\overline{A}_{\varepsilon }$ the trajectory tree
associated with $\{\overline{C}_{\varepsilon }^{(n)}\}$,
\begin{eqnarray*}
P^{(n)}(\overline{C}_{\varepsilon }^{(n)})\geq P(\overline{A}_{\varepsilon
})&=&P(A_{\varepsilon /2}\backslash \bigcup_{n\geq \overline{N}(\varepsilon )}%
\widetilde{A}_{\varepsilon /2}(n))\\ &>&1-\varepsilon /2-\varepsilon
/2=1-\varepsilon .
\end{eqnarray*}
Thus (4) is established. (1) and the $\varepsilon $-AEP are clearly
fulfilled, and (2) follows from the AEP as in step 1. Now (B) follows
immediately.$\qquad \qed$
\section{A quantum Shannon-McMillan-Breiman theorem}
We are now in a position to formulate and prove a quantum version of
Breiman's theorem. With this paper being a continuation of \cite{bksss1} we adopt
all settings from there, for simplicity  only considering the
one-dimensional case. That means that ${\cal A}^{\infty}$ is a
quasilocal $C^{*}$-algebra over $\zz$ constructed from a finite
dimensional $C^{*}$-algebra ${\cal A}$. For $m,n\in \zz$ with $m \leq
n$  we denote by ${\cal
  A}_{[m,n]}$ the local algebra over the
discrete interval $[m,n]\subset \zz$ and use the abbreviation ${\cal A}^{(n)}$ for ${\cal
  A}_{[1,n]}$, $n \in \nn$. Instead of ${\cal A}_{[n,n]}$ we write ${\cal A}_{[n]}$. Recall that for a stationary,
i.e. shift-invariant state $\Psi$ on ${\cal A}^{\infty}$ the mean entropy $s$ is defined as
\begin{eqnarray*}
s:= \lim_{n \to \infty}\frac{1}{n}S(\Psi^{(n)}),
\end{eqnarray*}
where $S(\Psi^{(n)}):=-\textrm{tr}D^{(n)}\log D^{(n)}$ is the von Neumann entropy of $\Psi^{(n)}$, the
restriction of the state $\Psi$ to the local algebra ${\cal
  A}^{(n)}$, and $D^{(n)}$ is the density operator from ${\cal A}^{(n)}$
corresponding to the state $\Psi^{(n)}$.\\
Let $A$ be a self-adjoint element of a local algebra ${\cal
  A}_{[m,n]}$. Then  $R(A)$ denotes its
range projector and $\textrm{tr}_{[k,l]}(A) $ is the partial trace of
$A$ over the
local algebra ${\cal A}_{[k,l]} \subset
{\cal A}_{[m,n]}$.

Observe that for the case that the underlying algebra
is abelian the following theorem reduces to the classical (reformulated)
Breiman assertion. So it can be considered as a quantum analogue of the SMB
theorem.
\begin{theorem}\label{breiman}
Let $\Psi $ be an ergodic state on the quasilocal C*-algebra $\mathcal{A}%
^{\infty }$ with mean entropy $s$. Then to each $\varepsilon >0$ there is a
sequence of orthogonal projectors $\{p_{\varepsilon }^{(n)}\}_{n=1}^{\infty }
$ in $\mathcal{A}^{(n)}$, respectively, and some $%
N(\varepsilon )$, such that
 \begin{enumerate}
  \item[ (q1)]\label{q1} $p_{\varepsilon }^{(n)}= R(\textup{tr}_{n+1
}\ (p_{\varepsilon }^{(n+1)}))$, \label{q1}

  \item[ (q2)] $\textup{tr }(p_{\varepsilon }^{(n)})\in (e^{n(s-\varepsilon
)},e^{n(s+\varepsilon )})$ for $n\geq N(\varepsilon )$, \label{q2}

  \item[ (q3)] there exist minimal projectors $p_{i} \in {\cal A}^{(n)}$   fulfilling $p_{\varepsilon }^{(n)}=\sum\limits_{i=1}^{%
\textup{tr }(p_{\varepsilon }^{(n)})}p_{i}$ and \(\Psi
^{(n)}(p_{i})<e^{-n(s-\varepsilon )}\) if $n\geq N(\varepsilon )$, \label{q3}

  \item[ (q4)] $\Psi ^{(n)}(p_{\varepsilon }^{(n)})>1-\varepsilon $. \label{q4}
 \end{enumerate}
\end{theorem}
\textbf{Remark:} It is a well known fact that each finite dimensional
unital $*$- algebra $ {\cal A} $ is isomorphic to
$\bigoplus_{i=1}^{s}{\cal B}(\hr_{i})$, where $\hr_{i}$ are finite
dimensional Hilbert spaces. According to this representation, we may
associate to each chained projector $p_{\eps}^{(n)}$ in the theorem
above a typical subspace of $\hr^{\otimes n}$ with $\hr:=\bigoplus_{i=1}^{s}\hr_{i}$. \\
\textbf{Proof of Theorem \ref{breiman}}\\
1. Let $\varepsilon >0$ be given. Choose an integer $l>0$ sufficiently
large such that the entropy of $\Psi^{(l)}$
satisfies $s\leq\frac{1}{l}S(\Psi
^{(l)})<s+\varepsilon ^{2}$. Take a complete set $V_{l}$ of mutually orthogonal spectral projectors for
$\Psi ^{(l)}$. Let $\mathcal{B}$ denote the abelian subalgebra of $%
\mathcal{A}^{(l)}$ generated by these projectors. The completeness of
$V_{l}$ implies that ${\cal B}$ is maximal abelian. Furthermore the
entropy of $\Psi^{(l)}\upharpoonright_{{\cal B}}$, the restriction of
$\Psi^{(l)}$ to the subalgebra ${\cal B}$, is identical to
$S(\Psi^{(l)})$. Generally we have the relation
\begin{eqnarray}\label{ua_entropy}
S(\Psi^{(n)})=\min\{S(\Psi^{(n)}\upharpoonright_{{\cal C}})|\ {\cal
  C}\subset {\cal A}^{(n)}\ \textsl{max. abelian subalgebra} \}.
\end{eqnarray}
The quasilocal algebra $\mathcal{B}^{\infty }$ constructed from ${\cal B}$
is an abelian subalgebra of $\mathcal{A}^{\infty
} $ and $\Psi $ acts on this algebra as a stochastic process $P_{l}$ with
alphabet $V_{l}$. The Shannon mean entropy $h_{l}$ of this process can
be estimated by $s\leq \frac{1}{l}h_{l} \leq
\frac{1}{l}S(\Psi^{(l)}\upharpoonright_{{\cal B}})<s+\varepsilon
^{2}$. The first inequality is a consequence of (\ref{ua_entropy}). $P_{l}$ is a
stationary, but not necessarily ergodic process. We apply the corresponding
version of the classical Shannon-McMillan-Breiman theorem (cf. \cite{keller},
\cite{Billingsley}) to this process and obtain that there is a set of trajectories $%
V_{l}^{\ast }\subset $ $V_{l}^{\mathbb{Z}}$ of measure one such that for
each $(v_{i})_{i \in \zz}\in V_{l}^{\ast }$ the limit
(individual mean entropy) $h((v_{i})_{i \in \zz
}):=\lim_{n\rightarrow \infty }-\frac{1}{n}\log
P_{l}^{(n)}((v_{i})_{i=1}^{n})$ exists, and we have $\mathbb{E}%
h((v_{i})_{i \in \zz})=h_{l}$.

2. Let $V_{l}^{\varepsilon ,-}\subset V_{l}^{\ast }$ be the subset of those
trajectories, for which the relation $\frac{1}{l}h((v_{i})_{i \in \zz})<s-\varepsilon ^{2}$ holds. We have $P_{l}(V_{l}^{\varepsilon
,-})=0.$ In fact, consider the sets
$W_{l}^{(n),-}:=\{(w_{i})_{i=1}^{n} \in
V^{(n)}_{l}|P_{l}^{(n)}((w_{i})_{i=1}^{n})>e^{-nl(s-\eps^{2})}\}$
obviously containing the sets $V_{l}^{(n),-}:=\{(v_{i})_{i=1}^{n} |\ \exists
(w_{i})_{i \in \zz}\in V_{l}^{\eps,-} \textsl{ with }
(w_{i})_{i=1}^{n}=(v_{i})_{i=1}^{n}\  \textsl{and }
P^{(m)}_{l}((v_{i})_{i=1}^{m})>e^{-ml(s-\eps^{2})}\ \textsl{for all }
m\geq n
\}$, respectively. This means that $P^{(n)}_{l}(W_{l}^{(n),-})\geq P^{(n)}_{l}(V_{l}^{(n),-})$. The
 cardinality of each $W_{l}^{(n),-}$
is bounded from  above by $e^{nl(s-\eps ^{2})}$.
Now suppose
$P_{l}(V_{l}^{\eps,-})>0$.
Then for $n$ sufficiently large we would have
$P_{l}^{(n)}(V_{l}^{(n),-})>c$ for some $c>0$ implying $P_{l}^{(n)}(W_{l}^{(n),-})>c.$
For the quantum state this would have the consequence that for large
$n$ there
are projectors $p^{(nl)}:=\sum_{(w_{i})_{i=1}^{n}\in
W_{l}^{(n),-}} \otimes_{i=1}^{n}w_{i}$ with $\textrm{tr }(p^{(nl)})<e^{nl(s-\varepsilon ^{2})}$ and $\Psi
^{(nl)}(p^{(nl)})>c$
. This contradicts Proposition 2.1 in \cite{bksss1}, saying that no
sequence of projectors in $\mathcal{A}^{(nl)}$, respectively, of
significant expectation can have asymptotically a smaller trace than $e^{nl(s-\varepsilon
^{2})}$.

3. Let $V_{l}^{\varepsilon ,+}\subset V_{l}^{\ast }$ be the subset of those
trajectories, for which the relation $\frac{1}{l}h((v_{i})_{i \in \zz})>s+\varepsilon $ holds. By 2. and by the relation $\mathbb{E}%
h((v_{i})_{i \in \zz})=h_{l}$ we obtain
\begin{eqnarray*}h_{l}>l(s-\varepsilon
^{2})(1-P_{l}(V_{l}^{\varepsilon ,+}))+l(s+\varepsilon
)P_{l}(V_{l}^{\varepsilon ,+})
\end{eqnarray*}
resulting in $P_{l}(V_{l}^{\varepsilon
,+})<2\varepsilon $.

4. Combining the preceeding results we can easily derive for each $%
\varepsilon >0$ the existence of an $l$ and of some $N(\varepsilon )$ such
that there is a subset $\widetilde{V}_{l}^{\ast }\subset V_{l}^{\ast }$ with
the properties
\begin{enumerate}
\item[(a)] $ P_{l}(\widetilde{V}_{l}^{\ast }) > 1-\varepsilon$ ,

\item[(b)] $e^{-nl(s+\varepsilon )}
< P_{l}^{(n)}((v_{i})_{i=1}^{n})<e^{-nl(s-\varepsilon )}$
for each
$(v_{i})_{i \in \zz}\in \widetilde{V}_{l}^{\ast }$\\  and
$n>N(\varepsilon )$.
\end{enumerate}
Indeed, assume $\eps <1$ (otherwise we would obtain the result above
with $\eps^{2}$ instead of $\eps$) and set $A_{l,\eps}:= V_{l}^{\ast
}\setminus (V_{l}^{\frac{\varepsilon}{2} ,-}\cup
V_{l}^{\frac{\varepsilon}{2} ,+})$.
We have $ P_{l}(A_{l,\eps})>1-\eps $ and $ A_{l,\eps}\subseteq
\bigcup_{n\geq 0}\bigcap_{k\geq n}A_{l,\eps}^{(k)}$, where
\[A_{l,\eps}^{(k)}:=\left\{(v_{i})_{i\in \nn}|\ -\frac{1}{k}\log
P_{l}^{(k)}((v_{i})_{i=1}^{k})\in (l(s-\eps),l(s+\eps))\right\}.\]
Then there exists $N(\eps)\in \nn$ such that $P_{l}(\bigcap_{k\geq
  N(\eps)}A_{l,\eps}^{(k)})>1-\eps $. The set $\widetilde{V}_{l}^{\ast
}:= \bigcap_{k\geq N(\eps)}A_{l,\eps}^{(k)}$ fulfills both conditions above.\\
Obviously $\widetilde{V}_{l}^{\ast }$ generates a sequence of chained sets $%
\{C_{\varepsilon }^{(n)}\}_{n=1}^{\infty }$ fulfilling \ref{(1)}-\ref{(4)} given in
section \ref{sect2}. In the given situation, we may reformulate these properties as
follows:
\begin{enumerate}
\item[(a)] $C_{\varepsilon }^{(n)}=(C_{\varepsilon }^{(n+1)})_{\flat }$ \ for $%
n\geq 1$ \

\item[(b)]$\;\#C_{\varepsilon }^{(n)}\in (e^{nl(s-\varepsilon
)},e^{nl(s+\varepsilon )})$ for $n\geq N(\varepsilon )$

\item[(c)] $\Psi ^{(nl)}(p\mathbf{)}<e^{-nl(s-\varepsilon )}$ for $n\geq
N(\varepsilon )$ and any $p=\otimes_{k=1}^{n}v_{k}$, where $%
(v_{k})_{k=1}^{n}$ $\mathbf{\in }C_{\varepsilon }^{(n)}$

\item[(d)]$\;\Psi ^{(nl)}(\sum_{(v_{k})_{k=1}^{n}\mathbf{\in }C_{\varepsilon
}^{(n)}}\otimes_{k=1}^{n}v_{k})>1-\varepsilon $ .
\end{enumerate}
Now it is easy to define chained projectors, first for multiples of $l$:
\begin{eqnarray*}
p_{\varepsilon }^{(nl)}:=\sum_{(v_{k})_{k=1}^{n}\mathbf{\in }C_{\varepsilon
}^{(n)}}\otimes_{k=1}^{n}v_{k},
\end{eqnarray*}
and then for general $n=ml+r,r<l$ by the set-up
\begin{eqnarray*}
p_{\varepsilon }^{(n)}:=R\left(\textrm{tr}_{[ml+r+1,(m+1)l]}\left(p_{\varepsilon
}^{((m+1)l)}\right)\right).
\end{eqnarray*}
Observe that both definitions are compatible. Obviously, by definition the
property (q1) is fulfilled by the defined system of projectors. Next, we
have with $n=ml+r,r<l$
\begin{eqnarray}\label{stern}
e^{n(s-2\varepsilon )}<\#C_{\varepsilon }^{(m)}\leq
\textrm{tr}(p_{\varepsilon }^{(n)}) &\leq& \#C_{\varepsilon }^{(m)}\textrm{tr}(
\mathbf{1}_{{\cal A}^{(l)}}\mathbf{)} \nonumber \\ &<& \textrm{tr}(\mathbf{1}_{
{\cal A}^{(l)}}\mathbf{)}e^{n(s+\varepsilon )}<e^{n(s+2\varepsilon
)}
\end{eqnarray}
for $n$ sufficiently large. In fact, the first inequality in this chain is obvious. By definition we have
\begin{eqnarray*}
p_{\varepsilon }^{(ml)}=\sum_{i=1}^{\textrm{tr}(p_{\varepsilon
}^{(ml)})}q_{i}^{(m)}
\end{eqnarray*}
for certain minimal projectors $q_{i}^{(m)}$ from $(\mathcal{B}
_{(l)})^{(m)}$, and
\begin{eqnarray*}
p_{\varepsilon }^{((m+1)l)}=\sum_{i=1}^{\textrm{tr }(p_{\varepsilon
}^{(ml)})}\sum_{j=1}^{k_{i}}q_{i}^{(m)}\otimes q_{i,j}
\end{eqnarray*}
for some minimal projectors $q_{i,j}$
from $\mathcal{B}_{[m+1]}$. In order to simplify our notation let
\[ I(m,r):=[ml+r+1,(m+1)l].\]
We obtain
\begin{eqnarray*}
\textrm{tr}(p_{\eps}^{(ml+r)}) &=&
\textrm{tr}\left(R \left( \sum_{i=1}^{\textrm{tr}(p_{\eps
}^{(ml)})}\sum_{j=1}^{k_{i}}q_{i}^{(m)}\otimes
\textrm{tr}_{I(m,r)}q_{i,j }\right) \right) \\
&=&
\textrm{tr}\left( \sum_{i=1}^{\textrm{tr}(p_{\varepsilon
}^{(ml)})}R\left( q_{i}^{(m)}\otimes \sum_{j=1}^{k_{i}}\textrm{tr}_{I(m,r)}q_{i,j}\right) \right) \\
&=&\sum_{i=1}^{\textrm{tr}(p_{\varepsilon }^{(ml)})}\textrm{tr}%
\left( q_{i}^{(m)}\otimes R\left( \sum_{j=1}^{k_{i}}\textrm{tr}_{I(m,r)}q_{i,j}\right) \right) \\
&=&\sum_{i=1}^{\textrm{tr}(p_{\varepsilon }^{(ml)})}\textrm{tr}
\left( R\left( \sum_{j=1}^{k_{i}}\textrm{tr}_{I(m,r)}q_{i,j}\right) \right) \\
&\geq &\textrm{tr}(p_{\varepsilon }^{(ml)})=\#C_{\varepsilon }^{(m)}.
\end{eqnarray*}
Here in the second step we made use of the mutual orthogonality of the $%
q_{i}^{(m)}$. This proves the second inequality in (\ref{stern}). The third
inequality also immediately follows from the formula
\begin{eqnarray*}
\textrm{tr}(p_{\varepsilon }^{(ml+r)})=\sum_{i=1}^{\textrm{tr}(p_{\varepsilon }^{(ml)})}\textrm{tr}\left( R\left(
\sum_{j=1}^{k_{i}}\textrm{tr}_{I(m,r)}(q_{i,j})\right) \right) .
\end{eqnarray*}
So (q2) is fulfilled, too (with $2\varepsilon $ instead of $\varepsilon $).

By (c) we see that (q3) is fulfilled if $n$ is a multiple of $l$. In the
general case $n=ml+r$ observe that in the representation
\begin{eqnarray*}
p_{\varepsilon }^{(ml+r)}=\sum_{i=1}^{\textrm{tr}(p_{\varepsilon
}^{(ml)})}q_{i}^{(m)}\otimes R\left( \sum_{j=1}^{k_{i}}\textrm{tr}_{I(m,r)}(q_{i,j})\right)
\end{eqnarray*}
we sum over mutually orthogonal projectors
each of them fulfilling
\begin{eqnarray*}
& &\Psi ^{(ml+r)}\left( q_{i}^{(m)}\otimes R\left( \sum_{j=1}^{k_{i}}\textrm{%
tr}_{I(m,r)}(q_{i,j})\right) %
\right) \\
&&\leq \Psi ^{(ml+r)}\left( q_{i}^{(m)}\otimes \mathbf{1}_{I(m,r)}\right)  \\
&&=\Psi ^{(ml)}(q_{i}^{(m)})<e^{-ml(s-\varepsilon )}<e^{-n(s-2\varepsilon )}
\end{eqnarray*}
if $n$ is sufficiently large. Now (q3) follows easily, again with $%
2\varepsilon $ instead of $\varepsilon $.\\
Finally, we have
\begin{eqnarray*}
&&\Psi ^{(ml+r)}(p_{\varepsilon }^{(ml+r)}) \\
&=&\sum_{i=1}^{\textsl{tr } (p_{\varepsilon }^{(ml)})}\Psi ^{(ml+r)}%
\left( q_{i}^{(m)}\otimes R\left( \sum_{j=1}^{k_{i}}\textrm{tr}_{I(m,r)}(q_{i,j})\right) \right)  \\
&=&\sum_{i=1}^{\textsl{tr }(p_{\varepsilon }^{(ml)})}\Psi ^{((m+1)l)}%
\left( q_{i}^{(m)}\otimes R\left( \sum_{j=1}^{k_{i}}\textrm{tr}_{I(m,r)}(q_{i,j})\right) \otimes \mathbf{1}_{%
\mathcal{A}_{I(m,r)}}\right)  \\
&\geq &\sum_{i=1}^{\textsl{tr }(p_{\varepsilon }^{(ml)})}\Psi
^{((m+1)l)}\left( q_{i}^{(m)}\otimes \sum_{j=1}^{k_{i}}q_{i,j}\right)
=\Psi ^{((m+1)l)}(p_{\varepsilon }^{((m+1)l)})>1-\eps,
\end{eqnarray*}
where we used the Schmidt representation and (d) to obtain the last line. This proves
(q4).$\qquad \qed $

\begin{acknowledgement}
We are deeply grateful to Ruedi Seiler for his continuous 
support of our work on a quantum version of the
Shannon-McMillan-Breiman Theorem and for many helpful discussions.\\
This work was supported by the DFG via the SFB 288 ``Differentialgeometrie und Quantenphysik'' at the TU Berlin and the Forschergruppe
``Stochastische Analysis und gro\ss e Abweichungen'' at University of Bielefeld.
\end{acknowledgement}

\end{document}